# A coherently excited Franson-type nonlocal correlation

Byoung S. Ham

School of Electrical Engineering and Computer Science, Gwangju Institute of Science and Technology, 123 Chumdangwagi-ro, Buk-gu, Gwangju 61005, S. Korea

(Submitted on April 27, 2023; bham@gist.ac.kr)

**Abstract**

Entanglement is the basic building block of quantum technologies whose property is in the unique quantum feature of nonlocal realism. However, such a nonlocal quantum property is known as just a weird phenomenon that cannot be obtained by any classical means. Recently, the mysterious quantum phenomena have been coherently interpreted using entangled photon pairs, where the quantum mystery has been found in the manipulated product-basis superposition of paired photons. Here, a coherence version of the Franson-type nonlocal correlation is presented by all means of classical physics. The resulting coherence solutions of the nonlocal correlation satisfy the same joint-phase relation of local parameters as in the quantum version. For the nonlocal correlation fringe, coherent manipulations of attenuated laser light are conducted by synchronized acousto-optic modulators to generate random but phase-matched photon pairs.

**Introduction**

Nonlocal quantum correlation between space-like separated two parties is a weird quantum phenomenon because it violates local realism [1,2]. Over the last several decades, various experimental demonstrations of the Bell inequality violation and delayed-choice quantum eraser have been conducted to understand the nonlocal quantum feature [3-10]. However, a complete understanding of the nonlocal correlation has been severely limited due to the local randomness-based joint-phase relation between independently controlled local parameters. Recently, a coherence approach [11-13] has been adopted to understand such weirdness of entangled photons generated from a spontaneous parametric down-conversion (SPDC) process [14,15] in nonlinear optics [16]. Like a typical phase-basis superposition of a single photon in an interferometric system, the coincidence detection-based selective measurement induces the second-order quantum superposition between measurement-selected tensor products of the paired photons [13].

Here, we propose a coherence model of Franson-type nonlocal correlation using a commercial laser to challenge the common understanding that a quantum feature cannot be achieved by any classical means. Thus, this paper is to revisit quantum entanglement to understand the role of quantum measurements such as in Franson correlation [8-11] and quantum erasers [12]. To mimic the same phase-matching relation as in the SPDC-generated entangled photons, a pair of synchronized acousto-optic modulators (AOMs) is used to coherently manipulate a narrowband laser light for a frequency-correlated photon pair. To satisfy incoherence conditions of the AOM-generated photon ensemble, a usual unbalanced Mark-Zehnder interferometer (UMZI) is used [13]. For this, the modulation bandwidth of the AOMs is set to be much wider than the laser linewidth. To satisfy detuning randomness, the AOM's scan speed is adjusted to be faster than the temporal resolution of single photon detectors. As a result, a coherence solution of the coincidence measurement-based two-photon correlation is obtained for the typical joint-phase relation of local parameters, satisfying the same Franson-type nonlocal correlation fringes.

**A coherence model of the Franson-type nonlocal correlation**

Figure 1 shows the proposed coherence model of the Franson-type nonlocal correlation using a commercial laser. To satisfy the statistical ensemble of photons, the laser is attenuated by neutral density filters [17]. For the phase-matching between paired coherent photons from UMZIs, the laser lights split by a beam splitter (BS1) are synchronously modulated by AOMs in an opposite frequency scanning mode (see the dotted boxes). For the UMZI, the path-length difference $\delta L$ between the long (L) and short (S) arms is set to be much longer than the effective coherence length determined by the AOM's modulation bandwidth $\Delta$ (see the Inset): $\delta L \gg c\Delta^{-1}$ and c is the speed of light. Local intensities of UMZI output photons should be uniform regardless of φ and ψ due to the incoherence condition for the photon ensemble. However, $\delta L$ is set to be shorter than the laser's coherence length $l_c$ to satisfy the single photon MZI physics [13]: $\delta L \ll l_c$. The photodetector's temporal resolution must cover $\Delta$. The AOM's frequency scan speed is set to be faster than the photo detector's resolving time to satisfy random choices of modulated photons in coincidence measurements. As a result, the paired coherent photons satisfy $\pm \delta f_j$ phase-matching relation.

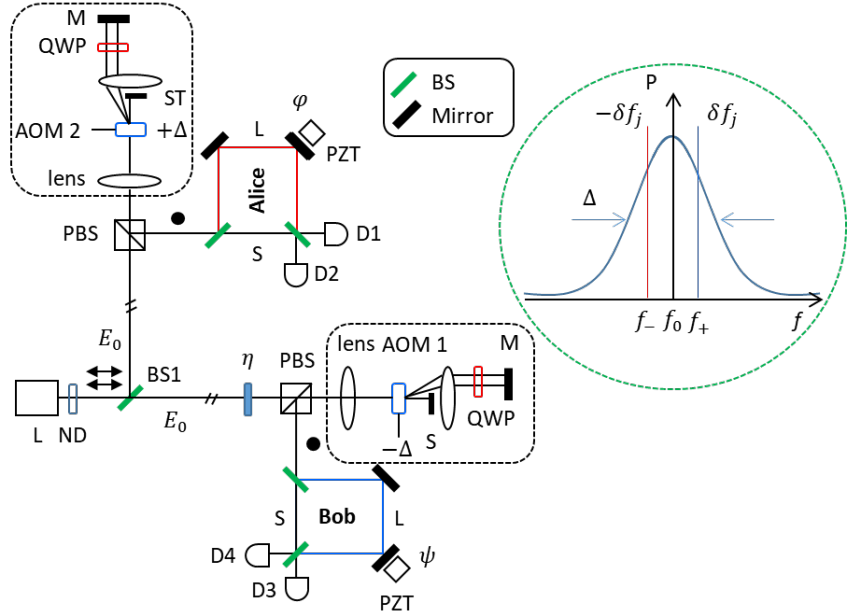

**Fig. 1. Schematic of macroscopic Franson-type nonlocal correlation.** AOM: acousto-optic modulator, BS: beam splitter, D: single photon detector, L: laser, ND: neutral density filter, QWP: quarter-wave plate, ST: beam stopper, PBS: polarizing BS, PZT: piezo-electric transducer. Inset: AOM pair-generated frequency pair. The $\pm \delta f_j$ represents a pair of AOM-generated photons at ~1 % ratio to the single photons [17]. The Gaussian-like probability envelope of the AOM-generated photon pairs is due to $\delta f_j$-dependent AOM diffraction efficiency.

The black dots in Fig. 1 indicate a vertically polarized photon pair statistically given by Poisson distribution, whereas the arrows represent a horizontally polarized photon pair. These photon pairs are of course post-determined by coincidence measurements [17]. As in the SPDC-generated entangled photons [14,15], coincidently measured photon pairs from UMZIs satisfy an opposite frequency relation $f_0 \pm \delta f_j$ by the AOM modulations. For a local phase control of the paired photons, a pair of piezo-electric transducers (PZTs) is independently manipulated for phase shifts $\varphi$ and $\psi$, respectively. Due to the coherence condition $l_c \gg \delta L$, each UMZI works as an interferometer to individual measurements. This individual measurement-based coherence condition is critical to understand the nonlocal intensity products via a coincidence detection technique (discussed in Discussion) [13]. If the AOM's scan speed is faster than the resolving time of the single-photon detector, the measured individual events should satisfy a general scope of a statistical ensemble of independent and individual events.

For the nonlocal correlation, quantum superposition between two paths of each UMZI is an essential requirement to excite the second-order amplitude superposition between measurement-selected tensor products (analyzed below). As discussed [13], the second-order amplitude superposition is caused by selective measurements of the coincidence detection applied only for the selected path-basis products, $\widehat{S_A S_B}$ and $\widehat{L_A L_B}$ (see Analysis). If orthogonally polarized photons are used in both UMZIs by replacing the BSs with polarizing BSs, the superposition between $\widehat{S_A S_B}$ and $\widehat{L_A L_B}$ product bases never be possible due to the impossible joint phase relation of $\varphi$ and $\psi$ (see Analysis).

**Analysis**

In both UMZIs of Fig. 1, the amplitudes of the $j^{th}$ output photon pair detected by D1-D4 are coherently represented by:

$$E_{1j}^A = \frac{E_0}{2} e^{i(k_0 r - \omega_0 t)} e^{\pm i(\delta k_j r - \delta \omega_j t)} \left(\widehat{S_A} - \widehat{L_A} e^{i(\xi_j \pm \varphi_j)}\right), \quad (1)$$

$$E_{2j}^A = \frac{iE_0}{2} e^{i(k_0 r - \omega_0 t)} e^{\pm i(\delta k_j r - \delta \omega_j t)} \left(\widehat{S_A} + \widehat{L_A} e^{i(\xi_j \pm \varphi_j)}\right), \quad (2)$$

$$E_{3j}^B = \frac{E_0}{2} e^{i\eta_j} e^{i(k_0 r - \omega_0 t)} e^{\mp i(\delta k_j r - \delta \omega_j t)} \left(\widehat{S_B} - \widehat{L_B} e^{-i(\xi_j \mp \psi_j)}\right), \quad (3)$$

$$E_{4j}^B = \frac{iE_0}{2} e^{i\eta_j} e^{i(k_0 r - \omega_0 t)} e^{\mp i(\delta k_j r - \delta \omega_j t)} \left(\widehat{S_B} + \widehat{L_B} e^{-i(\xi_j \mp \psi_j)}\right), \quad (4)$$

where A and B represent Alice and Bob, respectively. $\eta_j$ is a relative path-length caused phase between BS1 and the UMZIs, and $\xi_j$ is the given UMZI path-length difference $\delta L$-caused phase shift. The opposite signs between $\varphi_j$ (Alice) and $\psi_j$ (Bob) are due to the same PZT scanning direction, where $\xi_j = \frac{\delta f_j \delta L}{c}$, $\varphi_j = \frac{\delta f_j \delta L'}{c}$, $\psi_j = \frac{\delta f_j \delta L''}{c}$, $\delta L = L - S$, $\delta L'$ ($\delta L''$) is by Alice's (Bob's) PZT. Here, the coherent analysis is for classical physics based on Maxwell's equations of electromagnetic fields. By cavity optics, the coherence length (time) of the laser is the same as that of a single photon [17]. The unit vectors of the photon paths in UMZIs are represented by $\widehat{S_{A,B}}$ and $\widehat{L_{A,B}}$. $E_0$ is the amplitude of a single photon. Thus, the corresponding intensities measured in both UMZI output ports are as follows for an arbitrary $j^{th}$ photon:

$$I_{1j}^A = \frac{I_0}{2}\left(1 - \cos(\varphi_j')\right), \quad (5)$$

$$I_{2j}^A = \frac{I_0}{2}\left(1 + \cos(\varphi_j')\right), \quad (6)$$

$$I_{3j}^B = \frac{I_0}{2}\left(1 - \cos(\psi_j')\right), \quad (7)$$

$$I_{4j}^B = \frac{I_0}{2}\left(1 + \cos(\psi_j')\right), \quad (8)$$

where $I_0 = E_0 E_0^*$, $\varphi_j' = \xi_j \pm \varphi_j$, and $\psi_j' = \xi_j \mp \psi_j$. The cosine terms is based on UMZI coherence to each photon, satisfying indistinguishable photon characteristics. If the BS is replaced by a PBS, the cosine terms are vanished, satisfying distinguishable photon characteristics. Thus, the mean values of Eqs. (5)-(8) show a PZT-independent uniform intensity: $\langle I_1^A \rangle = \langle I_2^A \rangle = \langle I_1^B \rangle = \langle I_1^B \rangle = \frac{\langle I_0 \rangle}{2}$ due to $\langle \cos \varphi_j' \rangle = \langle \cos \psi_j' \rangle = 0$ for $\xi_j \gg 1$ ($\delta L \gg c\Delta^{-1}$), resulting in the ensemble incoherence-based local randomness [3-13].

For the nonlocal intensity correlation $R_{12}(\tau_{AB})$ between two local detectors from both UMZIs, e.g. D1 and D3, the coincidence detection selectively chooses $\widehat{S_A}\widehat{S_B}$ and $\widehat{L_A}\widehat{L_B}$ product bases only:

$$[R_{13}^{AB}(\tau_{AB})]_j = \frac{I_0^2}{16}\left(\widehat{S_A} - \widehat{L_A} e^{i(\xi_j \pm \varphi_j)}\right)\left(\widehat{S_B} - \widehat{L_B} e^{-i(\xi_j \mp \psi_j)}\right)(cc)$$

$$= \frac{I_0^2}{16}\left(\widehat{S_A}\widehat{S_B} + \widehat{L_A}\widehat{L_B} e^{\pm i(\varphi_j + \psi_j)}\right)(cc), \quad (9)$$

where cc is a complex conjugate. Here, the coincidence-detection time delay $\tau_{AB}$ is in the order of $\Delta^{-1}$, allowing sensitive variations by the PZT scan for $\varphi_j$ and $\psi_j$ in the order of the wavelength of the light. Thus, the mean value of Eq. (9) is as follows:

$$\langle R_{13}^{AB}(\tau_{AB} \sim 0) \rangle = \frac{\langle I_0^2 \rangle}{8}\langle 1 + \cos(\varphi_j + \psi_j) \rangle. \quad (10)$$

In Eq. (10), the cosine term shows the quantum feature of the joint-phase relation between independently controlled PZTs [18]. The direct local intensity product of Eqs. (5) and (7) cannot result in the joint-phase relation, prohibiting the quantum feature. Thus, the coherently excited quantum feature in Fig. 1 is due to coincidence detection-caused superposition of selective product bases for the AOM-manipulated coherent photon pairs. By the coincidence detection, the dropping noncoincidence terms $\widehat{S_A L_B} e^{i\psi_j}$ and $\widehat{S_B L_A} e^{i\varphi_j}$ plays a critical role in Eq. (9). Compared with $\widehat{S_A L_A}$ ($\widehat{S_B L_B}$) in Eqs. (5)-(8), $\widehat{S_A S_B L_A L_B}$ in Eq. (10) represents the second-order amplitude superposition between paired photons. Thus, the individual photon coherence in UMZIs is an essential requirement for Fig. 1.

Unlike Eqs. (5)-(8), Eq. (10) shows a nearly perfect phase correlation if $\tau_{AB} \ll \Delta^{-1}$, resulting in $\varphi_j + \psi_j \sim \varphi + \psi$ [8]. As $\tau_{AB}$ increases, the ensemble decoherence becomes dominant, resulting in decreased fringe visibility [19,20]. For $\tau_{AB} \gg \Delta^{-1}$, Eq. (10) shows the classical lower bound without fringes. Thus, the nonlocal quantum feature of the joint-phase relation is coherently derived for Fig. 1 using classical physics. In Eqs. (9) and (10), a path-length difference-based relative phase between $S_A$ and $S_B$ is neglected, otherwise, the joint phase $\psi + \varphi$ is simply shifted that much.

The mean intensity correlation between D2 and D4 is the same as Eq. (10) (not shown). Similarly, the mean intensity correlation between D1 and D4 (D2 and D3) is represented by:

$$\langle R_{14}^{AB}(\tau_{AB}\sim 0)\rangle = \langle R_{23}^{AB}(\tau_{AB}\sim 0)\rangle = \frac{\langle I_0{}^2\rangle}{8}\langle 1 - \cos(\varphi_j + \psi_j)\rangle. \quad (11)$$

In Eqs. (10) and (11), the opposite fringe patterns result from the MZI physics caused by BS. Thus, the coherence solutions of both local randomness and nonlocal intensity fringes are successfully derived for Fig. 1. The coherence condition of individual single photons under ensemble incoherence in UMZIs is the bedrock of the coherence interpretation for the nonlocal quantum feature.

**Discussion**

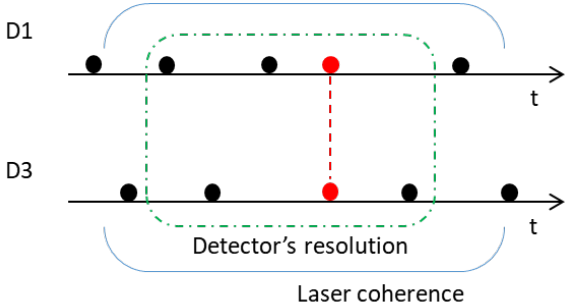

**Fig. 2. Schematic of photon streams from both UMZIs.** The red-dot pair indicates the coincidence detection resulting in the superposition relation between $\widehat{S_A S_B}$ and $\widehat{L_A L_B}$ product bases.

For the uniform local intensities derived in Eqs. (5)-(8), either a single photon or a coherent photon stream of an attenuated laser light gives the same MZI results due to the self-interference of a single photon [17,21]. However, the intensity-product correlations between two local detectors derived in Eqs. (10) and (11) do not give the same result in both cases due to the temporal resolution of a coincident photon pair (red dots) from other noncoincident single photons, as shown in Fig. 2. For a photodetector having a limited temporal resolution, crosstalk between a coincident photon (red dot) and non-coincident single photons (black dots) also contributes to the intensity products due to their longer coherence time. As analyzed in Eq. (9), a choice of path bases of a photon in both UMZIs for the coincidence detection is the key to the product-basis superposition between $\widehat{S_A S_B}$ and $\widehat{L_A L_B}$ with definite $\varphi_j$ and $\psi_j$ controls. This process is accomplished by the separation of coincident photon pairs (red dots) from others (black dots). Here, different time-slot photons must have different detuning $\delta f_j$, ($j \neq j'$), resulting in different $\varphi_j'$ and $\psi_j'$. One way to get rid of the crosstalk is to keep the mean photon number down to a single-photon resolvable level [17], which is the same as the single-photon detection scheme in conventional quantum measurements. In that sense, implementation of macroscopic nonlocal correlation using coherent light may not be possible. However, a mesoscopic pulsed scheme of Fig. 1 could be possible at the cost of visibility loss.

**Conclusion**

A coherence model of Franson-type nonlocal correlation was proposed and analyzed for both local randomness and nonlocal correlation fringes using classical physics. To generate phase-matched coherent photon pairs, a synchronized AOM pair was used to manipulate attenuated laser light whose modulation frequencies are opposite. The coherence solutions of the phase-matched coincident photon pairs for two-photon intensity correlation satisfied a typical joint-phase relation between independently controlled local (UMZI) parameters. The local randomness originated in the ensemble decoherence of individually coherent photons for UMZI. The secret of the usual coincidence detection technique applied to the Franson nonlocal correlation was selective choices of product bases between paired photons. This measurement modification was the origin of the product-basis superposition, resulting in the nonlocal quantum feature with a joint-phase relation of local parameters. Because this joint-phase relation is nonlocal, the AOM-manipulated photon pairs should be quantum, i.e., an entangled pair. Thus, the proposed coherently excited Franson-type nonlocal correlation was successfully analyzed by all means of classical physics. In a Poisson-distributed coherent photon stream, however, the ensemble effect of local self-interference implied a critical problem of inseparability between coincident (bunched) photons and other single photons. A possible solution to avoid such cross-correlation between paired photons and single photons was to resolve every single photon individually and independently. Although this photon-resolving solution in a spatial (temporal) domain was exactly for the conventional single-photon detection scheme, the proposed coherence motel was based on classical electromagnetic waves. Thus, the bedrock of the product-basis superposition was the phase correlation between the space-like separated UMZI output photons provided by AOM modulations.

**Reference**

1. Einstein, A., Podolsky, B. & Rosen, N. Can quantum-mechanical description of physical reality be considered complete? *Phys. Rev*. **47**, 777-780 (1935).
2. Bell, J. On the Einstein Podolsky Rosen Paradox. *Physics* **1**, 195-290 (1964).
3. Brunner, N. et al. Bell nonlocality. *Rev. Mod. Phys*. **86**, 419–478 (2014).
4. I. Marcikic, H. de Riedmatten, W. Tittel, H. Zbinden, M. Legre, and N. Gisin, Distribution of time-bin entangled qubits over 50 km of optical fiber. *Phys. Rev. Lett*. **93**, 180502 (2004).
5. X.-S. Ma, A. Qarry, J. Kofler, T. Jennewein, and A. Zeilinger. Experimental violation of a Bell inequality with two different degree of freedom of entangled particle pairs. *Phys. Rev. A* **79**, 042101 (2009).
6. B. Hensen et al., Loophole-free Bell inequality violation using electron spins separated by 1.3 kilometres. *Nature* **526**, 682-686 (2015).
7. Clauser, J. F., Horne, M. A., Shimony, A. & Holt, R. A. Proposed experiment to test local hidden-variable theories. *Phys. Rev. Lett*. **23**, 880–884 (1969).
8. Kwiat, P. G., Steinberg, A. M. & Chiao, R. Y. High-visibility interference in a Bell-inequality experiment for energy and time. *Phys. Rev. A* **47**, R2472–R2475 (1993).
9. Aerts, S., Kwiat, P., Larsson, J.-Å. & Żukowski, M. Two-photon Franson-type experiments and local realism. *Phys. Rev. Lett*. **83**, 2872-2875 (1999).
10. A. Cabello, A. Rossi, G. Vallone, F. De Martini, and P. Mataloni, Proposed Bell experiment with genuine energy-time entanglement. *Phys. Rev. Lett*. **102**, 040401 (2009).
11. G. Carvacho et al., Postselection-loophole-free Bell test over an installed optical fiber network. *Phys. Rev. Lett*. **115**, 030503 (2015).

12. B. S. Ham, A coherence interpretation of nonlocal realism in the delayed-choice quantum eraser. arXiv:2302.13474v2 (2023).
13. B. S. Ham, The origin of Franson-type nonlocal correlation. arXiv:2112.10148v4 (2023).
14. H. Cruz-Ramirez, R. Ramirez-Alarcon, M. Corona, and K. Garay-Palmett, and A. B. U'Ren, Spontaneous parametric processes in modern optics. *Opt. Photon. News* **22**, 36-41 (2011), and reference therein.
15. Zhang, C., Huang, Y.-F., Liu, B.-H, Li, C.-F. & Guo, G.-C. spontaneous parametric down-conversion sources for multiphoton experiments. *Adv. Quantum Tech.* **4**, 2000132 (2021).
16. Boyd, R. W. Nonlinear Optics. (Academic Press, New York, 1992), Ch. 2.
17. S. Kim and B. S. Ham, Revisiting self-interference in Young's double-slit experiments. *Sci. Rep.* **13**, 977 (2023).
18. Aerts, S., Kwiat, P. G., Larsson, J.-Å. &Zukowski, M. Two-photon Franson-type experiments and local realism. *Phys. Rev. Lett.* **83**, 2872-2875 (1999).
19. Herzog, T. J., Kwiat, P.G., Weinfurter, H. &Zeilinger, A. Complementarity and the quantum eraser. *Phys. Rev. Lett.* **75**, 3034-3037 (1995).
20. Riedinger, R., Wallucks, A., Marinković, I., Löschnauer, C., Aspelmeyer, M., Hong, S. & Gröblacher, S. Remote quantum entanglement between two micromechanical oscillators. *Nature* 10.1038/s41586-018-0036-z (2018).
21. Grangier, P., Roger, G. & Aspect, A. Experimental evidence for a photon Anti-correlation effect on a beam splitter: A new light on single-photon interferences. *Europhys. Lett.* **1**, 173–179 (1986).

**Author contribution**

B.S.H. solely wrote the manuscript. Correspondence and requests for materials should be addressed to BSH (email: bham@gist.ac.kr).

**Conflict of interest**

The author has no conflicts to disclose.

**Data availability**

Data sharing is not applicable – no new data is generated.

**Funding**

This work was supported by the ICT R&D program of MSIT/IITP (2023-2022-2021-0-01810), via Development of Elemental Technologies for Ultra-secure Quantum Internet. BSH also acknowledges that this work was supported by GIST-GRI 2023.